\documentclass[letterpaper, 10 pt, conference]{ieeeconf}  

\IEEEoverridecommandlockouts                              

\usepackage{amsmath} 
\usepackage{graphicx} 
\usepackage{amssymb}  
\usepackage{xcolor}
\usepackage{booktabs} 
\usepackage{multirow}
\usepackage{algorithm}
\usepackage{algpseudocode}
\usepackage{url}

\title{\LARGE \bf
Online Constraint Optimization-Based Indirect-Adaptive Robust Control for Autonomous Vehicles with Unknown Friction
}

\author{Yupeng 
Zhou$^{1}$, I-Chia Chang$^{1}$, Ziran Wang$^{1}$ and Bin Yao$^{1}$
\thanks{$^{1}$ College of Engineering, Purdue University, West Lafayette, IN 47906, USA
        {\tt\small byao@purdue.edu}}%
}

\begin{document}

\maketitle
\thispagestyle{empty}
\pagestyle{empty}

\begin{abstract}

\begingroup\color{black}
Friction uncertainty is a significant challenge for autonomous vehicle control, undermining feasibility and performance. This paper proposes a novel two-layer control framework to achieve vehicle motion control under large friction uncertainty by combining optimization-based planning with indirect adaptive robust control (IARC). The upper layer solves a constrained optimal control problem
to generate a feasible trajectory that satisfies all the constraints in a time-optimal manner along the desired path. The lower layer, IARC, adapts the unknown parameters related to friction uncertainty and combines model compensation with robust feedback to ensure transient and steady-state tracking performance. An estimation monitor is designed to prevent poor learning when data indicate a model mismatch, while a smooth parameter reset reduces the estimated tire capacity and limits abrupt tightening of constraints. The updated parameters are fed back into the model and constraints in the planner, enabling trajectory replanning as friction conditions change. Simulation results indicate that the proposed method achieves accurate tracking of the desired trajectory and online friction adaptation under large friction uncertainty while ensuring time-optimal vehicle motion. The simulation results demos are available at \url{https://github.com/TedZhou0603/CO-IARC}.
\endgroup

\end{abstract}

\section{INTRODUCTION}
Optimal control problems (OCPs), in particular, Nonlinear Model Predictive
Control (NMPC), are widely studied for autonomous vehicle motion control by incorporating nonlinear vehicle dynamics and
constraints~\cite{9790832}. NMPC applications range from lane keeping
and speed tracking~\cite{song2015combined} to demanding maneuvers such as
autonomous racing~\cite{9366415}~\cite{9340640}~\cite{9341731}~\cite{10645032}, obstacle avoidance~\cite{10752593,9115712}, autonomous drifting~\cite{goh2020toward}~\cite{goh2024beyond} and precise tracking~\cite{zarrouki2024r2nmpc}~\cite{10801876}. To achieve demanding maneuvers, the pivot of these studies is a high-fidelity
vehicle model, particularly with different sophisticated tire models that capture nonlinear wheel forces using accurate tire-road
friction information.

Physics-based tire models relate tire forces to slip, load, and road friction~\cite{goh2024beyond}, with parameters fitted from prior measurements~\cite{9366415}. Alternatively, data-driven models use neural networks to represent tire forces~\cite{djeumou2023autonomousdrifting3minutes} or vehicle dynamics~\cite{spielberg2019neural}, requiring training data that adequately cover relevant operating conditions~\cite{11146591}. When embedded in an NMPC, both approaches may lose accuracy if additional uncertainties, unmodeled terms, or tire properties differ from those represented in the offline identification or training data. This limitation is particularly important under low-friction conditions, where even routine maneuvers can approach tire-force limits. Laurense et al.~\cite{7963824} reported that the conventional tracking approach evaluated in their experiments required friction-coefficient accuracy within approximately $2\%$ at the friction limit, highlighting the importance of accurate friction information. However, repeated friction measurements can be costly and time-consuming, while collecting training data covering every possible road and tire condition is impractical.

Online adaptation for friction uncertainty addresses this limitation by updating the model online during operation. Wurts et al.~\cite{9147713} and Thompson et al.~\cite{10752593} used
an unscented Kalman filter (UKF) to estimate tire-road friction online and
update the prediction model in NMPC. However, the estimation accuracy and closed-loop performance do not establish convergence to the true physical parameters or explicit tracking-error bounds during adaptation for transient performance. Vaskov et
al.~\cite{9867523} learned lateral tire-friction characteristics using Gaussian-process models and incorporated estimation uncertainty into stochastic NMPC. Their probabilistic constraint considered uncertainty during learning, but does not establish an explicit deterministic bound for the unknown friction. As the vehicle is in operation while the friction estimates are not precise, the remaining challenge is still the lack of a guarantee for the transient and steady-state tracking performance.

To preserve online adaptation capability and maintain high-accuracy tracking under friction uncertainty in both transient and steady-state conditions, we propose a two-layer control framework inspired by the linear motor control approach~\cite{10551497,liu2025time}.
In our vehicle motion control design, the upper-layer OCP pursues
time-optimal vehicle motion to track the desired trajectory within the lane by encouraging
forward progress while accounting for vehicle dynamics and constraints.
The lower-layer indirect adaptive robust control (IARC)~\cite{yao1997adaptive} calculates model compensation with estimated tire-model parameters
from measurements collected during driving, without prior
road-friction measurements or offline training data.
In our design, the high-frequency robust feedback provides transient and steady-state longitudinal
speed-tracking error bounds without requiring parameter convergence~\cite{liu2025time}.

To properly account for vehicle and tire dynamics and maintain feasible control, it is necessary to identify tire-saturation conditions. Overestimating tire force capacity can create demands beyond the available grip and drive the tires into saturation. This will further lead the vehicle into an open-loop, unstable drifting regime, in which sideslip regulation and a separate model capturing saturated tire behavior become necessary~\cite{goh2020toward}. For normal driving conditions, the control design relies on operation within the unsaturated tire region. Saturation therefore poses difficulties for both control and estimation. For the controller, it can require unattainable control force. For the estimator, saturated data no longer satisfy the regression model, so continued adaptation will lead to model mismatch and increase parameter adaptation error. Since the parameter estimates also update the tire model and friction constraints in the planner, monitoring tire saturation is necessary to prevent unreliable adaptation, which can further compromise planning feasibility.

In this work, our main contributions are as follows:
\begin{itemize}
    \item We propose a novel two-layer optimization-based time-optimal indirect adaptive robust control framework for autonomous vehicle
    motion control under friction uncertainty. The upper-layer OCP plans constrained vehicle motion, while the lower-layer IARC provides tracking with guaranteed transient and steady-state performance, incorporating explicit online adaptation without requiring parameter convergence.
    \item An estimation monitor is developed to prevent poor learning when data indicate a model mismatch. A smooth parameter reset reduces the estimated tire capacity while limiting abrupt tightening of constraints. The updated estimates are fed back into the planner's model and constraints, supporting replanning under friction uncertainty and reducing the risk of infeasibility.
    \item High-fidelity simulations verify the proposed IARC effectively provides online friction estimation and adaptation. The transient and steady-state errors are guaranteed to be within the designed bound of $L_e=0.05\ \mathrm{m/s}$, indicating precise speed-tracking performance.
    The parameter feedback mechanism incorporates estimation monitoring to ensure feasible planning and time optimality under varying uncertain friction conditions from $\mu\in[0.2,1.05]$, reducing time consumption by 24.4\% compared with the conservative baseline.
\end{itemize}

\section{Modeling and Problem Formulation}
\subsection{Vehicle Dynamics Model}
In this work,  a rear-drive single-track 
dynamic model is obtained by lumping the left and the right tracks into a single centered track. Following~\cite{9366415}, vehicle motion is expressed in Frenet coordinates, where $s$ denotes the distance along the path curvature, $e$ the lateral displacement, and $ \Delta\psi$ the heading error relative to the path tangent. With roll and pitch ignored, the model is shown below
\begin{equation}
\begin{aligned}
    \dot{s} &= \frac{U_x \cos\Delta\psi - U_y \sin\Delta\psi}{1 - \kappa(s)e}\\
    \dot{e} &= U_x \sin\Delta\psi + U_y \cos\Delta\psi \\
    \Delta\dot{\psi} &= r - \kappa(s)\dot{s}\\
    \dot{U}_x &= \frac{F_{xf} \cos\delta - F_{yf} \sin\delta + F_{xr} - C_dU_x^2-F_{rr}}{m} + r U_y \\
    \dot{U}_y &= \frac{F_{yf} \cos\delta + F_{xf} \sin\delta + F_{yr} }{m} - r U_x\\
    \dot{r} &= \frac{a(F_{yf} \cos\delta + F_{xf} \sin\delta) - b F_{yr}}{I_{z}}
\label{eq:dynamics}
\end{aligned}
\end{equation}
\begingroup\color{black}
In \eqref{eq:dynamics}, $U_x$ and $U_y$ are the body-frame velocities
at the center of gravity, $r$ is the yaw rate, and $\kappa(s)$ is
the path curvature. The vehicle mass and yaw inertia are $m$ and
$I_z$, while $a$ and $b$ are the distances from the center of gravity
to the front and rear axles. For $i\in\{f,r\}$, $F_{xi}$ and $F_{yi}$
denote the longitudinal and lateral axle forces in the tire frame.
The steering angle is $\delta$, $C_dU_x^2$ is the aerodynamic drag
force, and $F_{rr}$ is the rolling resistance. Only rear longitudinal
actuation is considered, so $F_{xf}=0$. The model is used for forward
motion with $U_x>0$ and $1-\kappa(s)e>0$. For the upper-layer OCP, $\delta$ and
$F_{xr}$ are included as additional states, with $\dot\delta$ and
$\dot F_{xr}$ as inputs.
\par\endgroup
For simplicity, the lower-layer IARC focuses on longitudinal dynamics for speed tracking with a single-input single-output (SISO) adaptive robust
control design, while the upper-layer OCP retains the coupled
longitudinal, lateral, and yaw dynamics for coordinated motion planning. 
Neglecting the fast dynamics of wheel rotation gives
\begin{equation}
    \dot{U}_x = \frac{F_{xr} - C_dU_x^2-F_{rr}}{m} + r U_y + \Delta
    \label{eq:2}
\end{equation}
The measured coupling $rU_y$ is retained. The front lateral-force projection $-F_{yf}\sin\delta/m$, external disturbances, and remaining unmodeled effects are included in the lumped acceleration uncertainty $\Delta$.

\subsection{Tire Model}
\label{subsec:shared_tire_model}
In this paper, we focus on normal driving conditions rather than on drifting, in which the tire is unsaturated. The two layers use tire-force models suited to their respective tasks with the same parameter definition. For the high-level OCP, the longitudinal tire force is generated and determined by $\dot{F}_{x, i}$; the maximum available lateral force will be 
\begingroup\color{black}
\begin{equation}
\bar F_{y,i}=\sqrt{(\mu F_{z,i})^2-F_{x,i}^2},
\end{equation}
\endgroup
where $F_{z,i}$ is the normal tire load~\cite{9366415}. The tire-road friction is defined as $\mu$.
\textcolor{black}{The longitudinal force must satisfy $|F_{x,i}|\le\mu F_{z,i}$. The lateral force is}
\begingroup\color{black}
\begin{equation}
F_{y,i}=\begin{cases}
\begin{aligned}
&-C_i\tan\alpha_i
+\dfrac{C_i^2|\tan\alpha_i|\tan\alpha_i}{3\bar F_{y,i}}\\
&-\dfrac{C_i^3}{27\bar F_{y,i}^{\,2}}\tan^3\alpha_i,
\end{aligned}
&|\alpha_i|\le\alpha_{\mathrm{sl},i},\\[4pt]
-\bar F_{y,i}\operatorname{sgn}(\alpha_i),
&|\alpha_i|>\alpha_{\mathrm{sl},i},
\end{cases}
\label{eq:lateral-tire-force}
\end{equation}
\endgroup
where $C_i$ is the cornering stiffness, $\alpha_i$ is the lateral slip.
\textcolor{black}{Equation~\eqref{eq:lateral-tire-force} applies for $\bar F_{y,i}>0$,
with $\alpha_{\mathrm{sl},i}=\tan^{-1}(3\bar F_{y,i}/C_i)$;
at $\bar F_{y,i}=0$, $F_{y,i}=0$.}
\textcolor{black}{The} front and rear axle lateral slip
angles can be calculated by \textcolor{black}{$
\alpha_f
=
\tan^{-1}
\left(
\frac{U_y+ar}{U_x}
\right)
-\delta$}
and 
\textcolor{black}{$
\alpha_r
=
\tan^{-1}
\left(
\frac{U_y-br}{U_x}
\right)$}. 
For the longitudinal dynamics in \eqref{eq:2} for the IARC design, we use the unsaturated branch of the isotropic brush model in \cite{goh2024beyond}. Under the small-slip-angle approximation $\tan\alpha_r\approx\alpha_r$, the longitudinal tire force $F_{x, r}$ can be obtained by
\begingroup\color{black}
\begin{equation}
F_{x,r}=C_r\kappa_r
-\frac{C_r^2\kappa_r\sigma_r}{3\mu F_{z,r}}
+\frac{C_r^3\kappa_r\sigma_r^2}{27(\mu F_{z,r})^2}.
\end{equation}
\endgroup
\textcolor{black}{Here $\sigma_r=\sqrt{\alpha_r^2+\kappa_r^2}$, and the model applies in the unsaturated
region $\sigma_r<3\mu F_{z,r}/C_r$.}
The rear-wheel longitudinal slip is defined as $\kappa_r = \frac{R \omega_r - U_x}{U_x}$, where the rear wheel speed $\textcolor{black}{\omega_r}$ can be measured in real time. 
This unsaturated region is denoted by $\mathcal D=\{(\kappa_r,\alpha_r)\mid\sigma_r<3\mu F_{z,r}/C_r\}$.

\subsection{Problem Formulation}
With the vehicle and tire models introduced, we use {$\theta$} to denote the unknown parameter \textcolor{black}{vector} as $\theta =\begin{bmatrix}
    \theta_1,\theta_2
\end{bmatrix}^T = \begin{bmatrix}
\dfrac{1}{3\mu F_{z,r}},
\dfrac{1}{27(\mu F_{z,r})^2}
\end{bmatrix}^T$. \textcolor{black}{The unknown parameters depend on the rear
normal load $F_{z,r}$ and tire-road friction $\mu$. Although the exact values of these
parameters are unknown, the variations are limited. So we
made the following assumption.}

\textbf{Assumption:}
 The unknown parameter vector $\theta$ and the uncertain nonlinearity $\Delta$ are within a known bounded convex set
\begin{equation}
\theta \in \Omega_{\theta}
\triangleq
\left\{ \theta \mid \theta_{\min} < \theta < \theta_{\max} \right\}
\end{equation}

\begin{equation}
\Delta \in \Omega_\Delta
\triangleq
\left\{ \Delta \mid |\Delta| \leq \delta_\Delta \right\}
\end{equation}
where $\theta_{\max} = [\theta_{1\max}, \theta_{2\max}]^T$,
$\theta_{\min} = [\theta_{1\min}, \theta_{2\min}]^T$,
and $\delta_\Delta$ are known.

\textcolor{black}{Let $y_d$ denote the desired position trajectory. 
The objective is to
pursue time-optimal motion along $y_d$ with accurate tracking 
performance under bounded parametric uncertainty and nonlinearities, 
while ensuring the constraints in \eqref{eq:lg_nlp} are satisfied.}
\section{Control Design}
\subsection{\textcolor{black}{Control Framework}}
\begin{figure}[!t]
      \color{black}
      \centering
      \includegraphics[width=\linewidth]{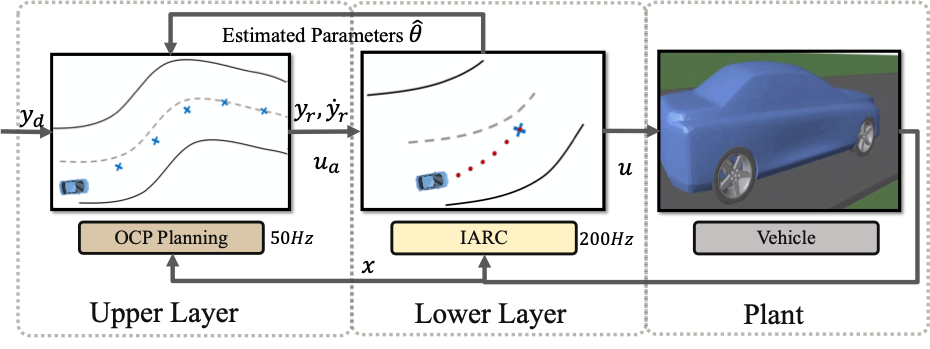}
      \caption{Two-layer Control Framework.}
      \label{fig:applied-inputs}
\end{figure}

\begingroup\color{black}
Fig.~\ref{fig:applied-inputs} illustrates the proposed two-layer control
framework for vehicle motion under friction uncertainty. The upper-layer
OCP operates at 50 Hz and uses updated parameter estimates
$\boldsymbol{\hat{\theta}}$ to generate feasible reference motion $y_r$
and $\dot{y}_r$ for tracking the desired trajectory $y_d$. The planner
pursues time-optimal motion while accounting for vehicle dynamics and
constraints. The resulting plan includes the longitudinal motion
reference and the steering.

The lower-layer IARC operates at 200 Hz to track the longitudinal speed
reference $U_{x,\mathrm{ref}}$ associated with the planned motion $y_r$
through a SISO design. The control input combines model
compensation $u_a$ with robust feedback to form the normalized
longitudinal force command $u$. The model compensation accounts for the
reference motion and known vehicle dynamics, while the robust feedback
corrects speed-tracking errors in the presence of parameter uncertainty
and unmodeled effects. The IARC tracking-error bound and feedback-force
reserve determine the planner constraint margins.

The measured vehicle state $x$ is fed back to both layers and supports
online estimation of the unknown parameters in the lower layer. The
updated estimates $\boldsymbol{\hat{\theta}}$ are fed back to the upper
layer to update the model and constraints to support time-efficient
and feasible replanning.
\par\endgroup
\subsection{Lower Layer Indirect-Adaptive Robust Control}
The lower-layer IARC includes online parameter estimation, estimator
monitoring, and control design. The monitor supervises the
data for the adaptation, while the controller uses the current parameter estimates
$\boldsymbol{\hat{\theta}}$ for speed tracking. The feedback force required for tracking is then bounded to determine the reserve for upper-layer
planning.
\subsubsection{\textbf{Adaptation Law}}
With the longitudinal vehicle
dynamics for speed tracking in (\ref{eq:2}), the unsaturated longitudinal tire force can be parameterized by 
\begingroup\color{black}
\begin{equation}
\begin{aligned}
F_{xr}
&=C_r\kappa_r+\varphi(\kappa_r,\alpha_r)^T\theta
\end{aligned}
\end{equation}
Here the regressor is $\varphi(\kappa_r,\alpha_r)=
\begin{bmatrix}
-C_r^2\kappa_r\sigma_r,C_r^3\kappa_r\sigma_r^2
\end{bmatrix}^T$.
Define
\begingroup\small
\begin{equation}
q=\frac{C_r\kappa_r-C_dU_x^2-F_{rr}}m+rU_y,
\quad \dot U_x=q+\frac{\varphi^T\theta}{m}+\Delta.
\label{eq:arc-identification-model}
\end{equation}
\endgroup
\endgroup
Use measured slip and the stable filter
$H_f(s)=\lambda_f/(s+\lambda_f)$, $\lambda_f>0$.
\textcolor{black}{Let $(\bullet)_f$ denote a filtered signal, $\hat{(\bullet)}$ denote estimate signals, and define the
filtered acceleration regressor as $\varphi_f=H_f[\varphi/m]$.}
To avoid differentiating the measured velocity, construct
\begin{equation}
\begin{aligned}
\dot U_{x,f}&=\lambda_f(U_x-U_{x,f}),
\quad U_{x,f}(0)=U_x(0),\\
y_f&=\lambda_f(U_x-U_{x,f})-q_f
=\varphi_f^T\theta+\Delta_f.
\end{aligned}
\label{eq:arc-filtered-regression}
\end{equation}
This initialization removes the derivative-filter initial
transient. With the predicted output $\textcolor{black}{\boldsymbol{\hat{y}}}_f=\varphi_f^T\textcolor{black}{\boldsymbol{\hat{\theta}}}$, the prediction error $\epsilon$ can be calculated by
\begin{equation}
\epsilon=y_f-\textcolor{black}{\boldsymbol{\hat{y}}}_f
=-\varphi_f^T\textcolor{black}{\boldsymbol{\tilde{\theta}}}+\Delta_f.
\label{eq:arc-prediction-error}
\end{equation}
where \(\textcolor{black}{\boldsymbol{\tilde{\theta}}}=\textcolor{black}{\boldsymbol{\hat{\theta}}}-\theta\). With this structured model, we used the normalized least-squares adaptation law proposed in \cite{yao2002indirect} 
\begin{equation}
\begin{aligned}
\textcolor{black}{\boldsymbol{\dot{\hat{\theta}}}}
&=
\operatorname{Proj}_{\textcolor{black}{\boldsymbol{\hat{\theta}}}}
\left(
    \Gamma\tau
\right),                                                   \\[2pt]
\dot{\Gamma}(t)
&=
\begin{cases}
\mu_1\Gamma
-
\frac{\Gamma\varphi_f\varphi_f^T\Gamma}
{1+\varphi_f^T\Gamma\varphi_f},
&
\text{if } \lambda_M\!\left(\Gamma\right)<\rho_M
\\
0,
&
\text{else}.
\end{cases}
\end{aligned}
\label{eq:arc-estimator-update}
\end{equation}
The $\operatorname{Proj}_{\textcolor{black}{\boldsymbol{\hat{\theta}}}}(\bullet)$ is a projection
operator defined as
\begingroup\color{black}
\begin{equation}
\begin{aligned}
    \operatorname{Proj}_{\textcolor{black}{\boldsymbol{\hat{\theta}}}}(\Gamma\tau)
    =
    \begin{cases}
        \Gamma\tau,
        & \begin{aligned}
          &\text{if } \textcolor{black}{\boldsymbol{\hat{\theta}}} \in \mathring{\Omega}_{\theta}\\
          &\text{or } n_{\textcolor{black}{\boldsymbol{\hat{\theta}}}}^{T}\Gamma\tau \leq 0,
          \end{aligned}\\[2pt]
        \displaystyle
        \left(
            I-\Gamma
            \frac{n_{\textcolor{black}{\boldsymbol{\hat{\theta}}}}n_{\textcolor{black}{\boldsymbol{\hat{\theta}}}}^{T}}
            {n_{\textcolor{black}{\boldsymbol{\hat{\theta}}}}^{T}\Gamma n_{\textcolor{black}{\boldsymbol{\hat{\theta}}}}}
        \right)\Gamma\tau,
        & \begin{aligned}
          &\textcolor{black}{\boldsymbol{\hat{\theta}}} \in \partial\Omega_{\theta}\\
          &\text{and } n_{\textcolor{black}{\boldsymbol{\hat{\theta}}}}^{T}\Gamma\tau > 0
          \end{aligned}
\end{cases}
\end{aligned}
\end{equation}
\endgroup
where $\Gamma$ is a symmetric positive definite matrix and $\tau=\frac{\varphi_f\textcolor{black}{\epsilon}}
{1+\varphi_f^T\Gamma\varphi_f}$
is the estimation function.
The sets $\partial\Omega_{\theta}$ and $\mathring{\Omega}_{\theta}$
denote the boundary and interior of $\Omega_{\theta}$, respectively,
and $n_{\textcolor{black}{\boldsymbol{\hat{\theta}}}}$ is the outward-pointing unit normal vector
at $\textcolor{black}{\boldsymbol{\hat{\theta}}} \in \partial\Omega_{\theta}$.
This projection operator ensures that
\begin{equation}
    \textcolor{black}{\boldsymbol{\hat{\theta}}}(t) \in \bar{\Omega}_{\theta}
    =
    \left\{
        \textcolor{black}{\boldsymbol{\hat{\theta}}}
        \mid
        \theta_{\min} \leq \textcolor{black}{\boldsymbol{\hat{\theta}}} \leq \theta_{\max}
    \right\},
    \quad \forall t
\end{equation}

\begingroup\color{black}
\subsubsection{\textbf{Estimator Monitoring Design}}
\textcolor{black}{The estimator uses the unsaturated tire model for normal
driving. During tire saturation, the measured force--slip response no
longer follows this model. Using these data to update the parameters can
produce incorrect estimates. The monitor therefore determines whether
the data can be used for learning.}

\textcolor{black}{Define the estimated local slope of rear longitudinal
force with respect to longitudinal slip as}
\begingroup\color{black}
\begin{equation}
\boldsymbol{\hat{S}}_k
\triangleq
\left[\boldsymbol{\frac{\partial \hat{F}_{xr}}{\partial\kappa_r}}\right]_k,
\label{eq:arc-monitor-slope}
\end{equation}
\endgroup
\textcolor{black}{where the derivative is estimated from the force--slip
data window ending at monitor update $k$. The threshold $c_\kappa>0$
specifies the minimum slope accepted for learning. Between monitor
updates, the nominal adaptation law is}
\begingroup\color{black}
\begin{equation}
\textcolor{black}{\boldsymbol{\dot{\hat{\theta}}}}=
\begin{cases}
\operatorname{Proj}_{\textcolor{black}{\boldsymbol{\hat{\theta}}}}(\Gamma\tau),
  & \boldsymbol{\hat{S}}_k\ge c_\kappa,\\
0, & \text{otherwise}.
\end{cases}
\label{eq:arc-slip-monitor}
\end{equation}
\endgroup

Algorithm~\ref{alg:slope_monitor} suspends the nominal parameter and
covariance updates when $\boldsymbol{\hat{S}}_k<c_\kappa$. The reset
$\Gamma\gets\Gamma_0$ restores the prescribed large initial adaptation
gain for faster parameter adjustment when learning resumes. The matrix
is held at $\Gamma_0$ until $\boldsymbol{\hat{S}}_k\ge c_\kappa$, when the
updates in~\eqref{eq:arc-estimator-update} resume.
\begin{algorithm}[H]
\color{black}
\caption{Estimator Monitoring}
\label{alg:slope_monitor}
\begin{algorithmic}[1]

\State Initialize $\boldsymbol{\hat{\theta}}\gets\boldsymbol{\hat{\theta}}_0$,
$\Gamma\gets\Gamma_0$, and the force--slip data window
\State Choose the smooth-reset factor $0<\gamma_r<1$

\For{each monitor update $k$}
    \State Update the force--slip data window
    \State Obtain the local slope $\boldsymbol{\hat{S}}_k$ from the window

    \If{$\boldsymbol{\hat{S}}_k<c_\kappa$}
        \Comment{adaptation off}
        \State $\boldsymbol{\hat{\theta}}
        \gets
        \gamma_r\boldsymbol{\hat{\theta}}
        +(1-\gamma_r)\theta_{\max}$
        \Comment{smooth reset}
        \State $\boldsymbol{\dot{\hat{\theta}}}\gets0$, $\Gamma\gets\Gamma_0$, and $\dot{\Gamma}\gets0$
    \Else
        \Comment{adaptation on}
        \State $\boldsymbol{\dot{\hat{\theta}}}
        \gets
        \operatorname{Proj}_{\boldsymbol{\hat{\theta}}}(\Gamma\tau)$
        \State Apply the covariance update in~\eqref{eq:arc-estimator-update}
    \EndIf

\EndFor

\end{algorithmic}
\end{algorithm}
While learning is suspended, the smooth reset moves
$\boldsymbol{\hat{\theta}}$ toward $\theta_{\max}$ at each monitor
update. An abrupt parameter change can tighten the OCP constraints so
that the fixed initial state $\boldsymbol{x}_0$ violates the updated hard
bounds, making the problem infeasible. The smooth reset reduces this
risk by changing the planning limits gradually. For $0<\gamma_r<1$, the
reset preserves the parameter bounds; smaller $\gamma_r$ gives a faster
correction.

\subsubsection{\textbf{Control Law Design}}
\begingroup\color{black}
Once $\boldsymbol{\hat{\theta}}$ is obtained by the adaptation law constructed above, let $z_1=U_x-U_{x,\mathrm{ref}}$ denote the speed tracking error.
Using $\textcolor{black}{\boldsymbol{\hat{\theta}}}$ from the estimator, define the normalized virtual
input and express the actual tire force as
\begin{equation}
u\triangleq\frac{C_r\kappa_r+\varphi^T\textcolor{black}{\boldsymbol{\hat{\theta}}}}{m},
\quad \frac{F_{xr}}m=u-\frac{\varphi^T\textcolor{black}{\boldsymbol{\tilde{\theta}}}}{m}.
\label{eq:arc-force-residual}
\end{equation}
The control design assumes that this virtual input is realizable
within the prescribed operating domain.
\endgroup
The reference is continuous and piecewise continuously
differentiable. To compensate the known dynamics using measured
states, construct the normalized force command as
\begin{equation}
\begin{aligned}
u&=u_a+u_{s1}+u_{s2},\\
u_a
&=
\dot U_{x,\mathrm{ref}}
+\frac{C_dU_x^2+F_{rr}}{m}
-rU_y,\\
u_{s1}&=-k_1z_1,
\qquad k_1>0.
\end{aligned}
\label{eq:arc-control-decomposition}
\end{equation}
Substituting \eqref{eq:arc-control-decomposition} into the true
speed dynamics therefore yields
\begingroup\color{black}
\begin{equation}
\begin{aligned}
\dot z_1
&=
u-\frac{C_dU_x^2+F_{rr}}m+rU_y
-\dot U_{x,\mathrm{ref}}
-\frac{\varphi^T\textcolor{black}{\boldsymbol{\tilde{\theta}}}}{m}+\Delta\\
&=
-k_1z_1+u_{s2}
-\frac{\varphi^T\textcolor{black}{\boldsymbol{\tilde{\theta}}}}{m}+\Delta.
\end{aligned}
\label{eq:arc-error-dynamics}
\end{equation}

Choose a known constant $h_1$ satisfying
$|-\varphi^T\textcolor{black}{\boldsymbol{\tilde{\theta}}}/m+\Delta|\le h_1$ throughout the
prescribed domain. For independent parameter intervals, a
sufficient choice is
\begin{equation}
\begin{aligned}
h_1\ge{}&
\frac{C_r^2\bar\kappa\bar\sigma}{m}
(\theta_{1\max}-\theta_{1\min})\\
&+
\frac{C_r^3\bar\kappa\bar\sigma^2}{m}
(\theta_{2\max}-\theta_{2\min})
+\delta_\Delta.
\end{aligned}
\label{eq:arc-robust-bound}
\end{equation}
Here $|\kappa_r|\le\bar\kappa$ and $\sigma_r\le\bar\sigma$.
\textcolor{black}{A smaller bound can replace \eqref{eq:arc-robust-bound} when
the coefficient errors are physically coupled.} In either case, $h_1$ must bound the
residual uniformly on the prescribed domain.
Select $u_{s2}$ to satisfy the ARC robust-performance conditions
\begin{equation}
\begin{aligned}
z_1u_{s2}&\le0,\\
z_1\bigl(
u_{s2}-\frac{\varphi^T\textcolor{black}{\boldsymbol{\tilde{\theta}}}}{m}+\Delta
\bigr)
&\le\varepsilon_1,
\qquad \varepsilon_1>0.
\end{aligned}
\label{eq:arc-robust-conditions}
\end{equation}
\endgroup
Following the bounded robust feedback in
\textcolor{black}{\cite{10551497}}, choose
\begin{equation}
u_{s2}
=
-h_1\operatorname{sat}
\left(
\frac{h_1z_1}{4\varepsilon_1}
\right),
\label{eq:arc-robust-feedback}
\end{equation}
where $\operatorname{sat}(s)=s$ for $|s|\le1$, and
$\operatorname{sat}(s)=\operatorname{sgn}(s)$ otherwise.
Its local gain is $h_1^2/(4\varepsilon_1)$, while
$|u_{s2}|\le h_1$.

\noindent\textit{Lemma 1:}
With \eqref{eq:arc-control-decomposition},
\textcolor{black}{\eqref{eq:arc-force-residual}}, and
\eqref{eq:arc-robust-feedback}, the speed tracking error satisfies
\begin{equation}
\begin{aligned}
z_1^2(t)
\le{}&
e^{-2k_1t}z_1^2(0)\\
&+
\frac{\varepsilon_1}{k_1}
\left(
1-e^{-2k_1t}
\right).
\end{aligned}
\label{eq:arc-tracking-bound}
\end{equation}

\noindent\textit{Proof:}
Let $V_1=z_1^2/2$. The first inequality in
\eqref{eq:arc-robust-conditions} follows directly from
\eqref{eq:arc-robust-feedback}. If $h_1|z_1|\le4\varepsilon_1$,
\begingroup\color{black}
\begin{equation}
\begin{aligned}
&z_1\bigl(
u_{s2}-\frac{\varphi^T\textcolor{black}{\boldsymbol{\tilde{\theta}}}}{m}+\Delta
\bigr)\\
&\qquad\le
-\frac{h_1^2z_1^2}{4\varepsilon_1}
+h_1|z_1|
\le\varepsilon_1.
\end{aligned}
\label{eq:arc-boundary-layer-proof}
\end{equation}
\endgroup
Otherwise, the same expression is bounded by
$-h_1|z_1|+h_1|z_1|=0$. Hence, using
\eqref{eq:arc-error-dynamics},
\begin{equation}
\dot V_1
\le
-k_1z_1^2+\varepsilon_1
=
-2k_1V_1+\varepsilon_1.
\label{eq:arc-lyapunov-inequality}
\end{equation}
The comparison lemma gives \eqref{eq:arc-tracking-bound}.

Consequently, the guaranteed ultimate error radius is
$\sqrt{\varepsilon_1/k_1}$. For a prescribed bound $L_e>0$,
the conditions
\begin{equation}
|z_1(0)|\le L_e,
\qquad
0<\varepsilon_1\le k_1L_e^2
\label{eq:arc-invariant-error-bound}
\end{equation}
\textcolor{black}{ensure $|z_1(t)|\le L_e$ on the prescribed domain, provided the
complete control command can be realized.}

\begingroup\color{black}
This tracking bound does not require persistent excitation (PE).
Parameter convergence requires the additional conditions below.

\noindent\textit{Lemma 2:}
Suppose $\theta\in\bar\Omega_\theta$ is constant and $\Delta=0$.
For $t\ge t_0$, suppose parameter adaptation remains active
without resets, $\mu_1\ge0$, $\Gamma$ is uniformly positive
definite and bounded, and $\varphi_f$ is bounded. If
\begin{equation}
\int_t^{t+T}\varphi_f(\nu)\varphi_f^T(\nu)\,d\nu
\ge\varepsilon_p I_2,
\qquad \forall t\ge t_0,
\label{eq:arc-parameter-pe}
\end{equation}
for some $T,\varepsilon_p>0$, where $I_2$ is the $2\times2$
identity matrix, then $\textcolor{black}{\boldsymbol{\tilde{\theta}}}(t)\to0$ as $t\to\infty$.

\noindent\textit{Proof:}
The result follows from the parameter-convergence argument in
\cite{1243991}, applied to the filtered regression
\eqref{eq:arc-filtered-regression} under the stated conditions.
\endgroup

\begingroup\color{black}
Following the input-reserve construction in \cite{10551497},
the feedback command for $|z_1|\le L_e$ satisfies
\begin{equation}
\begin{aligned}
|u_{s1}+u_{s2}|&\le u_{sM},\\
u_{sM}
&=k_1L_e+
\min\left\{
\frac{h_1^2L_e}{4\varepsilon_1},h_1
\right\}.
\end{aligned}
\label{eq:arc-feedback-reserve}
\end{equation}
Since $u$ is a force command divided by mass, the required
feedback force reserve is
\begin{equation}
F_{sM}=m u_{sM}.
\label{eq:arc-force-reserve}
\end{equation}
For attainable longitudinal force limits
$F_{xr,\min}$ and $F_{xr,\max}$, the feedforward command must satisfy
\begin{equation}
F_{xr,\min}+F_{sM}\le m u_a\le F_{xr,\max}-F_{sM}.
\label{eq:arc-feedforward-force-limits}
\end{equation}
These margins reserve feedback authority for both acceleration
and braking. Here $F_{sM}$ bounds the longitudinal feedback
correction $m(u_{s1}+u_{s2})$. The tightened interval must be nonempty.
For fixed $h_1$, $k_1$, $\varepsilon_1$, and $L_e$, $F_{sM}$ is
a constant bound, independent of the current parameter estimate.
\endgroup
 
\subsection{Upper Layer Optimization}
\begingroup\color{black}
At each planning update, the upper layer uses the current vehicle
state and the updated tire capacities $\textcolor{black}{\boldsymbol{\hat{\mu}}}\textcolor{black}{\boldsymbol{\hat{F}}}_{z,r}$ and
$\textcolor{black}{\boldsymbol{\hat{\mu}}}\textcolor{black}{\boldsymbol{\hat{F}}}_{z,f}$, obtained from the parameter estimates
$\textcolor{black}{\boldsymbol{\hat{\theta}}}$, to replan feasible vehicle motion. The finite-horizon
OCP is solved using acados~\cite{verschueren2020acadosmodularopensourceframework}.

Let $t_0,\ldots,t_N$ denote the prediction nodes over $N$
prescribed intervals $\Delta t_k=t_{k+1}-t_k>0$.
The planning states and inputs are
\begin{equation}
\begin{aligned}
\boldsymbol x_k
&=[s_k,e_k,\Delta\psi_k,U_{x,k},U_{y,k},r_k,\delta_k,F_{xr,k}]^T,\\
\boldsymbol u_k&=[\dot\delta_k,\dot F_{xr,k}]^T.
\end{aligned}
\label{eq:ocp-state-input}
\end{equation}

The OCP is formulated as
\endgroup
\begin{equation}
\begin{gathered}
 \textcolor{black}{\underset{\boldsymbol x_k,\boldsymbol u_k,\boldsymbol z_k}
 {\operatorname{minimize}}\quad
 J_{\mathrm{progress}}+J_{\mathrm{rd}}+J_{\mathrm{input}}+J_\alpha+J_F}
 \\[0.5em]
 \begin{alignedat}{2}
 &\text{s.t.}\\[-0.2em]
 &\boldsymbol x_0=\boldsymbol x_{\mathrm{initial}}
       &\quad&\text{(Dynamics)}\\
 &\boldsymbol x_{k+1}
     =\textcolor{black}{f_{\mathrm{st}}^{\Delta t_k}(\boldsymbol x_k,\boldsymbol u_k)}\\
 &\textcolor{black}{U_x^{\min}+L_e\leq U_{x,k}}
       &&\text{(State limits)}\\
 &\textcolor{black}{U_{x,k}\leq f_{env}(y_d,\textcolor{black}{\boldsymbol{\hat{\theta}}})-L_e}\\
 &\delta^{\min}\leq\delta_k\leq\delta^{\max}\\
 &\textcolor{black}{F_{xr,\min}+F_{sM}\leq F_{xr,k}}\\
 &\textcolor{black}{F_{xr,k}\leq F_{xr,\max}-F_{sM}}\\
 &F_{y}^{\min}\leq F_{y}\leq \textcolor{black}{\boldsymbol{\hat{F}}}_{y}^{max}(\textcolor{black}{\boldsymbol{\hat{\theta}}})\\
 &\textcolor{black}{\dot{F}_{xr}^{\min}\leq \dot{F}_{xr,k}\leq \dot{F}_{xr}^{\max}}
       &&\text{(Input limits)}\\
 &\dot\delta^{\min}\leq\dot\delta_k\leq\dot\delta^{\max}\\
 &e_k^{\min}\leq e_k\leq e_k^{\max}
       &&\text{(Road boundaries)}\\
 &{|\alpha_{i,k}|
    \leq 0.85 \alpha_{\mathrm{sl},i,k}(\boldsymbol{\hat{\theta}})+s_{\alpha_i,k}}
       &&\text{(Slip limits)}\\
 &\textcolor{black}{\sqrt{F_{xf,k}^{2}+F_{yf,k}^{2}}}
       &&\text{(Friction limits)}\\
 &\textcolor{black}{\qquad\leq{\boldsymbol{\hat{\mu}}}{\boldsymbol{\hat{F}}}_{z,f}+s_{F_f,k}}\\
 &\textcolor{black}{\sqrt{(|F_{xr,k}|+F_{sM})^{2}+F_{yr,k}^{2}}}\\
 &\textcolor{black}{\qquad\leq\textcolor{black}{\boldsymbol{\hat{\mu}}}\textcolor{black}{\boldsymbol{\hat{F}}}_{z,r}+s_{F_r,k}}\\
 \end{alignedat}
\end{gathered}
\label{eq:lg_nlp}
\end{equation}
\begingroup\color{black}
The dynamics $f_{\mathrm{st}}^{\Delta t_k}$ uses Runge--Kutta
integration with known road geometry and current tire-parameter
estimates. The initial condition fixes $\boldsymbol x_0$ to the
current state. $f_{\mathrm{env}}(\boldsymbol{y}_d,\textcolor{black}{\boldsymbol{\hat{\theta}}})$ is the upper-speed
envelope~\cite{yu2025spatialenvelopempchigh} to preprocess the reference
trajectory $\boldsymbol{y}_d$ based on estimated parameters
$\textcolor{black}{\boldsymbol{\hat{\theta}}}$. Tightening both speed limits by $L_e$ accommodates the IARC error bound in \eqref{eq:arc-invariant-error-bound}. Since the vehicle tracks the previous plan, which satisfies the tightened constraints, Lemma~1 ensures that $\boldsymbol x_0$ satisfies the original speed limits $[U_x^{\min},\,f_{\mathrm{env}}]$; the tightened constraints in \eqref{eq:lg_nlp} are therefore imposed for $k=1,\ldots,N$. The lower bound $U_x^{\min}$ and
allowance $L_e$ complete the speed interval, while
$[\delta^{\min},\delta^{\max}]$ represents the steering range.

The rear-force limits reserve $F_{sM}$ from
\eqref{eq:arc-force-reserve} for the ARC control input. 
Force-rate and steering-rate bounds limit command variations, while
\endgroup
$e_k\in[e_k^{\min},e_k^{\max}]$ defines lateral displacement to the
safe road corridor.
For each axle $i\in\{f,r\}$, the slip constraint
limits the magnitude of the lateral slip angle $\alpha_{i,k}$ to a prescribed
fraction of $\alpha_{\mathrm{sl},i,k}(\boldsymbol{\hat{\theta}})$. This limit is the
sliding slip angle $\alpha_{\mathrm{sl},i}$ in~\eqref{eq:lateral-tire-force},
evaluated using the predicted longitudinal force and estimated tire capacity.
\textcolor{black}{The nonnegative slack $s_{\alpha_i,k}$ allows relaxation
for additional model mismatch and too tight feasibility and is penalized through
$J_{\alpha}$ in the total objective $J$.}

\begingroup\color{black}
The objective rewards forward progress while penalizing
deviation from the reference path $\boldsymbol{y}_d$, rapid control changes, and slip- and friction-limit violations:
\begin{itemize}
    \item \textit{Forward progress:}
    $J_{\mathrm{progress}}=-W_p s_N$, with $W_p>0$, rewards
    greater travel distance over the prescribed prediction horizon.

    \item \textit{Reference deviation:}
    $J_{\mathrm{rd}}=W_e e_N^2+W_\psi(\Delta\psi_N)^2$
    penalizes terminal lateral displacement and heading error
    relative to the reference path $\boldsymbol{y}_d$.

    \item \textit{Control smoothness:}
    $J_{\mathrm{input}}$ sums weighted quadratic penalties on
    $\dot\delta_k$ and $\dot F_{xr,k}$ over $k=0,\ldots,N-1$,
    limiting abrupt steering and longitudinal-force changes.

    \item \textit{Slip penalty:}
    $J_\alpha$ sums weighted linear and quadratic penalties on
    the nonnegative slacks $s_{\alpha_i,k}$ for both
    axles over $k=0,\ldots,N$, discouraging violations of the
    prescribed slip limits.

    \item \textit{Friction penalty:}
    $J_F$ sums weighted linear and quadratic penalties on
    $s_{F_i,k}$ for both axles over $k=0,\ldots,N$,
    discouraging exceedance of the friction bounds.
\end{itemize}
All penalty weights are nonnegative, with positive weights
on the friction slacks.
\endgroup

\begingroup\color{black}
The planner supplies the speed reference, nominal rear-force
profile, and steering command.
Let $T_{\mathrm{IARC}}$ denote the IARC sampling period.
Within interval $k$, define $t_{k,i}=t_k+iT_{\mathrm{IARC}}$
for integers $i\geq 0$ satisfying
$iT_{\mathrm{IARC}}<\Delta t_k$.
At these updates, the reference signals are reconstructed as
\begin{equation}
\begin{aligned}
F_{xr,\mathrm{ref}}(t_{k,i})
&=F_{xr,k}+iT_{\mathrm{IARC}}\dot F_{xr,k},\\
U_{x,\mathrm{ref}}(t_{k,i})
&=U_{x,k}+\frac{iT_{\mathrm{IARC}}}{\Delta t_k}
  (U_{x,k+1}-U_{x,k}).
\end{aligned}
\label{eq:planner-arc-interpolation}
\end{equation}
The steering command is reconstructed from $\delta_k$ and
$\dot\delta_k$ in the same way as the rear force. The speed
interpolation gives
$\dot U_{x,\mathrm{ref}}=(U_{x,k+1}-U_{x,k})/\Delta t_k$.
At each fast update, the IARC evaluates the compensation force
$m u_a$ from \eqref{eq:arc-control-decomposition} using this
reference acceleration and the measured vehicle states, and
forms the force command $m u=m u_a+m(u_{s1}+u_{s2})$.
\endgroup

\section{Experiment}

\subsection{Experiment Setup}
\begingroup\color{black}
Simulations use Simscape Multibody Vehicle Dynamics with the vehicle configuration based on the test platform in~\cite{zhou2025hierarchicaltestplatformvision} with a $2411\ \mathrm{m}$ long test track provided in Monteblanco~\cite{10645032}. The straights and curves of the track require acceleration, braking, and cornering maneuvers typical of driving, testing the proposed method's ability to coordinate longitudinal and lateral tire-force demands within available friction limits. The vehicle starts on the track centerline at $12.44\ \mathrm{m/s}$.
The drivable lane width is set to $4.5\ \mathrm{m}$. 
 A $23\ \mathrm{m}$ launch section
with $\mu=1.0$ as dry road patch is followed by five sections with
$\mu=0.3$, $0.5$, $0.7$, $0.8$, and $0.4$, which consider the ice, wet, and ponding road friction patches.
Transitions between these different friction patches occur at approximately
$s=460$, $952$, $1326$, and $1899\ \mathrm{m}$; the last
section extends to the end of the lap, which covers different cornering, straight lines with deceleration and acceleration zones. The following four methods are evaluated on the same track and friction map:
\begin{itemize}
    \item[\textbf{M1:}] \textbf{Ours.}
    The proposed OCP--IARC framework with online parameter
    estimation, estimator monitoring, and parameter feedback
    to the planner. The initial friction estimate is
    $\hat{\mu}(0)=1.0$.
    \item[\textbf{M2:}] \textbf{DRY.}
    Nominal NMPC without IARC, with the prior friction
    coefficient fixed at $\mu=1.0$ throughout the run,
    representing a high-friction assumption for dry roads.
    \item[\textbf{M3:}] \textbf{WET.}
    The same nominal NMPC as M2, with the prior friction
    coefficient fixed at $\mu=0.6$, representing a
    moderate-friction assumption for wet roads.
    \item[\textbf{M4:}] \textbf{ICE.}
    The same nominal NMPC as M2, with the prior friction
    coefficient fixed at $\mu=0.3$, representing a
    low-friction assumption for icy roads.
\end{itemize}

\begin{table}[!t]
\color{black}
\setlength{\tabcolsep}{1.5pt}
\caption{Performance on Test Track with Different Friction Patches.}
\label{tab:tracking_comparison}
\begin{center}
\color{black}
\begin{tabular}{cccccccc}
\hline\hline
\multirow{2}{*}{Method} & \multirow{2}{*}{time/lap (s)}
  & \multicolumn{2}{c}{$e_\mathrm{lat}$ (m)}
  & \multicolumn{2}{c}{$e_\mathrm{speed}$ (mm/s)}
  & \multicolumn{2}{c}{$e_\mathrm{yaw}$ (deg)} \\
& & RMSE & max & RMSE & max & RMSE & max \\
\hline
DRY & \begin{tabular}[c]{@{}c@{}}crashed\\($s=200$ m)\end{tabular}
& $\geq 6$ & $\geq 6$ & 3.103 & 8.011 & 3.521 & 22.747 \\
WET & \begin{tabular}[c]{@{}c@{}}crashed\\($s=262$ m)\end{tabular}
& $\geq 6$ & $\geq 6$ & 2.161 & 11.721 & 6.254 & 21.021 \\
ICE & 202.7
& 0.213 & 0.779 & 0.203 & 0.735 & 1.418 & 5.467 \\
\textbf{Ours} & \textbf{153.2}
& \textbf{0.208} & \textbf{0.772}
& \textbf{0.168} & \textbf{0.688}
& \textbf{1.224} & \textbf{5.389} \\
\hline\hline
\end{tabular}
\end{center}
\end{table}

\subsection{Parameter Selection}
\begingroup\color{black}
The upper-level OCP is solved at 50 Hz, with a prediction horizon of 5.04 seconds divided into 40 intervals. The lower-level IARC runs at 200 Hz. The friction coefficient bound is set to $\mu\in[0.2,1.05]$,
with initial estimate $\textcolor{black}{\boldsymbol{\hat{\mu}}}(0)=1.0$. Accounting for load transfer
gives the rear tire force capacity interval
$\mu F_{z,r}\in[1592,12367]\ $. The corresponding
parameter bounds are
\mbox{$\theta_1\in[2.69\times10^{-5},\,2.79\times10^{-4}]$}
and
\mbox{$\theta_2\in[2.42\times10^{-10},\,2.60\times10^{-8}]$}.
The estimator is initialized with
$\Gamma(0)=\operatorname{diag}(5\times10^6,5\times10^8)$.
For the tracking-error allowance $L_e=0.05\ $,
the simulation uses $h_1=0.74$,
$k_1=h_1/(2L_e)=7.4$, and
$\varepsilon_1=k_1L_e^2\approx0.02$.
Equation~\eqref{eq:arc-feedback-reserve} gives the feedback force
reserve $F_{sM}=m u_{sM}\approx1385.5\ \mathrm{N}$.
\begin{figure}[thpb!]
      \color{black}
      \centering
      \includegraphics[width=\linewidth]{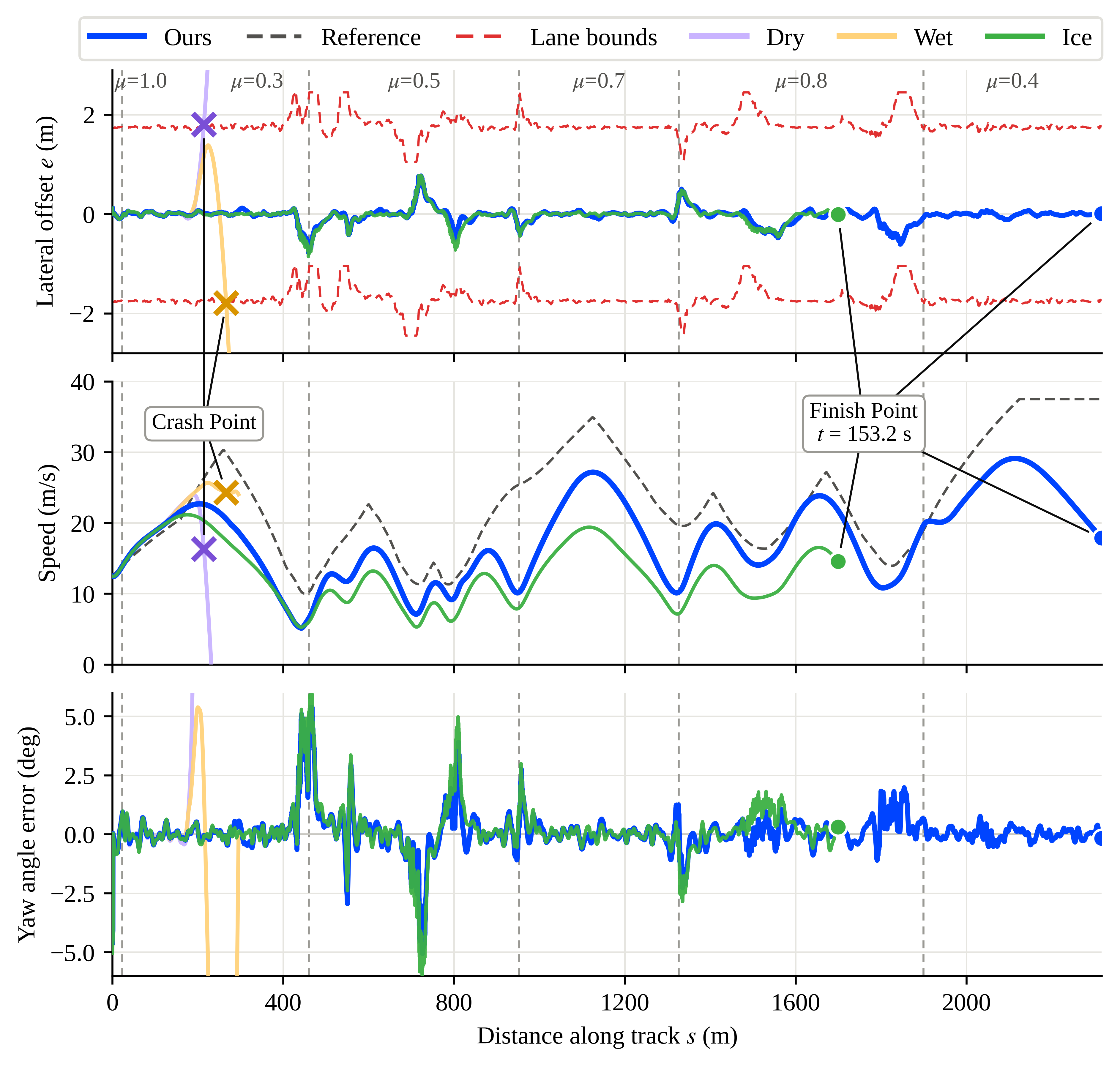}
    \caption{Lateral error, speed, and yaw-angle error for the proposed method and the baseline NMPC with different prior estimates of road friction. Our method achieves the best tracking performance and time optimality compared with a high-prior NMPC that crashed under low-friction conditions, and is faster than the most conservative prior.}
    \label{fig:tracking-comparison}
\end{figure}

\subsection{Results}
\begingroup\color{black}
Figure~\ref{fig:tracking-comparison} and
Table~\ref{tab:tracking_comparison} compare the proposed method
with nominal NMPC without IARC. With different priors
$\mu=1.0$ (DRY) and $0.6$ (WET), both nominal controllers overestimate
grip in the initial $\mu=0.3$ patch and crash at
$s=200$ and $262\ \mathrm{m}$, respectively. And maximum
lateral errors are at least $6\ \mathrm{m}$, with peak yaw errors
of $22.747^\circ$ and $21.021^\circ$. These failures show the
consequence of overestimated tire capacity.

With the lower prior $\mu=0.3$, ICE remains conservative after the grip increases. When the proposed method finishes the lap at $t=153.2$~s, ICE has covered only $1697$~m ($70.4\%$ of the lap), and it needs $202.7$~s to complete the lap, $32\%$ longer than the proposed method.

Our proposed method is able to complete the
full lap in $153.2\ \mathrm{s}$, with minimum lateral RMSE and maximum error of $0.208$ and
$0.772\ \mathrm{m}$, and its maximum speed error is
$0.688\ \mathrm{mm/s}$. Figure~\ref{fig:tracking-comparison}
shows close tracking of the replanned speed during acceleration
and braking, with no lane-bound violations. The two-layer design
achieves fast lap completion with accurate tracking
under changing friction, supporting the time-optimal planning
objective.
\begin{figure}[thpb!]
      \color{black}
      \centering
      \includegraphics[width=\linewidth]{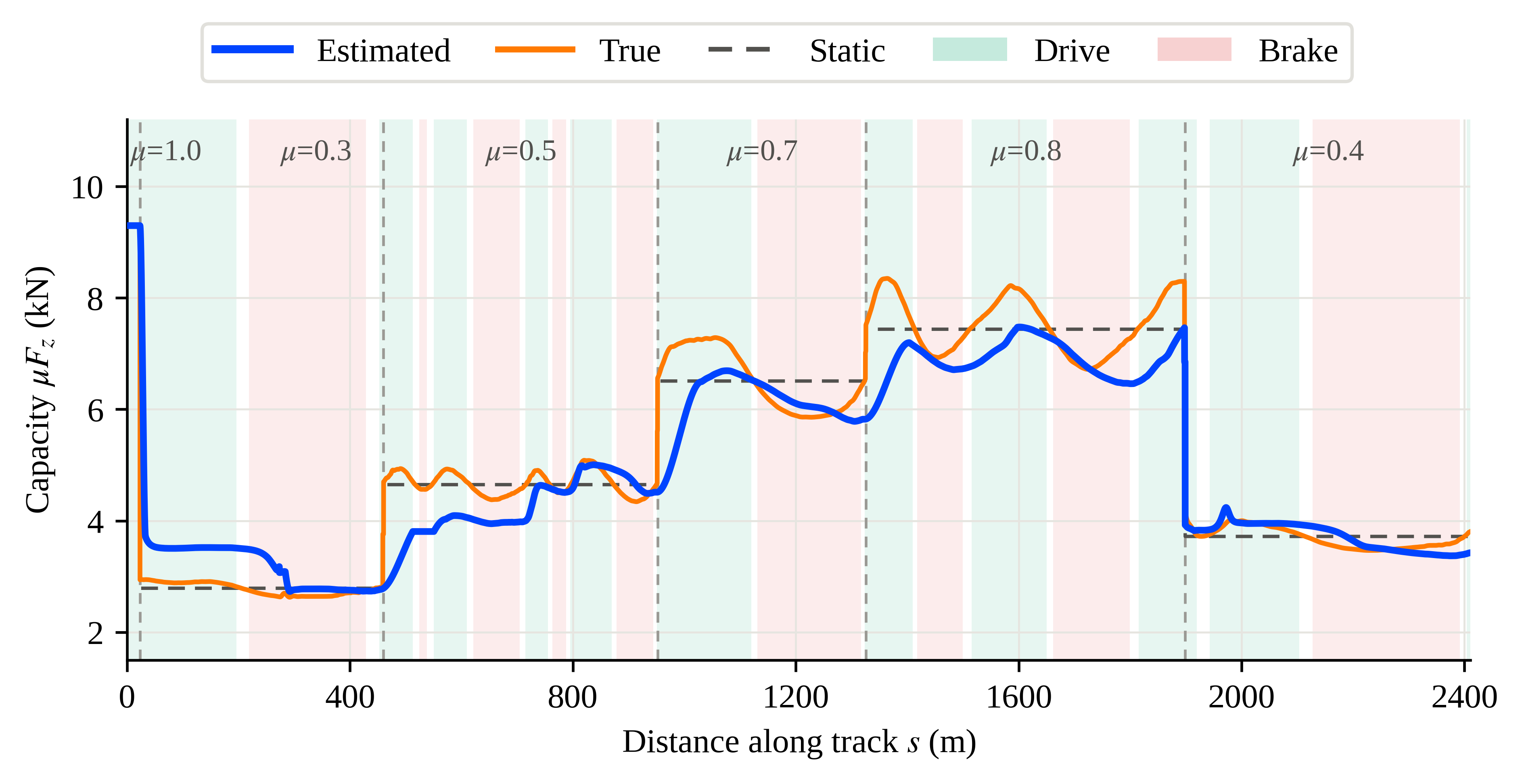}
      \caption{Estimated and true rear tire capacities. The gray dashed curve gives the capacity based on the static rear load. Green and pink shading indicate driving and braking; vertical dashed lines mark friction transitions.}
      \label{fig:friction-estimation}
\end{figure}
\textcolor{black}{Figure~\ref{fig:friction-estimation} shows the estimated rear tire
capacity adapting to changes in available grip. Friction changes and
load transfer make the true capacity time varying, while filtering and
finite adaptation response can introduce lag. At the transitions from
$\mu=1.0$ to $0.3$ and from $0.8$ to $0.4$, the estimated capacity
drops sharply, reducing the previous overestimate. These corrections
are consistent with the action of Algorithm~\ref{alg:slope_monitor}:
nominal learning is suspended when force--slip data indicate model
mismatch, and the smooth parameter reset reduces the tire capacity
used by the planner. In the higher-grip sections, the estimate increases
progressively during acceleration and braking. This recovery is
consistent with learning resuming on accepted data and allows the
planner to use the additional grip. Together with the
accurate speed tracking and lane-bound satisfaction in
Fig.~\ref{fig:tracking-comparison}, these responses support monitored
parameter feedback for feasible replanning under changing friction,
with the prescribed tracking allowance $L_e$ and feedback force
reserve $F_{sM}$.}
\endgroup
\section{CONCLUSIONS}
\begingroup\color{black}
This paper presented a two-layer OCP--IARC framework for autonomous vehicle motion control under friction uncertainty. The upper layer pursues time-optimal motion, while the lower layer provides explicit transient and steady-state longitudinal speed-tracking bounds under the stated assumptions, without requiring parameter convergence. Tracking-error and feedback-force margins are incorporated into the OCP constraints to account for the control effort required during adaptation.

To address tire saturation, an estimation monitor suspends nominal learning when force--slip measurements indicate model mismatch. A smooth parameter reset reduces estimated tire capacity while limiting abrupt constraint tightening. The monitored estimates update the planner's tire model and friction constraints, enabling online replanning without prior friction measurements or offline training data.

High-fidelity simulations on a circuit with multiple friction patches demonstrated full-lap completion in $153.2\ \mathrm{s}$ with a lateral tracking RMSE of $0.208\ \mathrm{m}$. The recorded longitudinal speed-tracking errors remained below the designed allowance of $L_e=0.05\ \mathrm{m/s}$. In comparison, fixed-friction NMPC with optimistic friction assumptions crashed in the low-friction section, whereas the conservative baseline needed $202.7$~s to complete the lap, $32\%$ longer than the proposed method. The proposed method adjusted its tire-capacity estimates during both grip reductions and recovery, supporting continued replanning and accurate tracking. These results demonstrate the value of coordinating monitored parameter adaptation with robust tracking and explicit planning margins under varying friction conditions.
\endgroup





\bibliography{reference}

@article{song2015combined,
  author={Song, Pan and Zong, Changfu and Tomizuka, Masayoshi},
  title={Combined Longitudinal and Lateral Control for Automated Lane Guidance of Full Drive-by-Wire Vehicles},
  journal={SAE International Journal of Passenger Cars - Electronic and Electrical Systems},
  volume={8},
  number={2},
  pages={419--424},
  year={2015},
  doi={10.4271/2015-01-0321}
}

@ARTICLE{9790832,
  author={Betz, Johannes and Zheng, Hongrui and Liniger, Alexander and Rosolia, Ugo and Karle, Phillip and Behl, Madhur and Krovi, Venkat and Mangharam, Rahul},
  journal={IEEE Open Journal of Intelligent Transportation Systems}, 
  title={Autonomous Vehicles on the Edge: A Survey on Autonomous Vehicle Racing}, 
  year={2022},
  volume={3},
  number={},
  pages={458-488},
  doi={10.1109/OJITS.2022.3181510}}

@ARTICLE{9366415,
  author={Laurense, Vincent A. and Gerdes, J. Christian},
  journal={IEEE Transactions on Control Systems Technology}, 
  title={Long-Horizon Vehicle Motion Planning and Control Through Serially Cascaded Model Complexity}, 
  year={2022},
  volume={30},
  number={1},
  pages={166-179},
  doi={10.1109/TCST.2021.3056315}}

@INPROCEEDINGS{9340640,
  author={Pagot, Edoardo and Piccinini, Mattia and Biral, Francesco},
  booktitle={2020 IEEE/RSJ International Conference on Intelligent Robots and Systems (IROS)}, 
  title={Real-time optimal control of an autonomous RC car with minimum-time maneuvers and a novel kineto-dynamical model}, 
  year={2020},
  volume={},
  number={},
  pages={2390-2396},
  doi={10.1109/IROS45743.2020.9340640}}

@INPROCEEDINGS{9341731,
  author={Vázquez, José L. and Brühlmeier, Marius and Liniger, Alexander and Rupenyan, Alisa and Lygeros, John},
  booktitle={2020 IEEE/RSJ International Conference on Intelligent Robots and Systems (IROS)}, 
  title={Optimization-Based Hierarchical Motion Planning for Autonomous Racing}, 
  year={2020},
  volume={},
  number={},
  pages={2397-2403},
  doi={10.1109/IROS45743.2020.9341731}}

@INPROCEEDINGS{10645032,
  author={Zarrouki, Baha and Wang, Chenyang and Betz, Johannes},
  booktitle={2024 American Control Conference (ACC)}, 
  title={A Stochastic Nonlinear Model Predictive Control with an Uncertainty Propagation Horizon for Autonomous Vehicle Motion Control}, 
  year={2024},
  volume={},
  number={},
  pages={5466-5473},
  doi={10.23919/ACC60939.2024.10645032}}

@ARTICLE{10752593,
  author={Thompson, Michael and Dallas, James and Goh, Jonathan Y. M. and Balachandran, Avinash},
  journal={IEEE Transactions on Field Robotics}, 
  title={Adaptive Nonlinear Model Predictive Control: Maximizing Tire Force and Obstacle Avoidance in Autonomous Vehicles}, 
  year={2024},
  volume={1},
  number={},
  pages={318-331},
  doi={10.1109/TFR.2024.3497925}}

@ARTICLE{9115712,
  author={Wurts, John and Stein, Jeffrey L. and Ersal, Tulga},
  journal={IEEE Transactions on Vehicular Technology}, 
  title={Collision Imminent Steering at High Speed Using Nonlinear Model Predictive Control}, 
  year={2020},
  volume={69},
  number={8},
  pages={8278-8289},
  doi={10.1109/TVT.2020.2999612}}

@article{goh2020toward,
  title={Toward automated vehicle control beyond the stability limits: drifting along a general path},
  author={Goh, Jonathan Y and Goel, Tushar and Christian Gerdes, J},
  journal={Journal of Dynamic Systems, Measurement, and Control},
  volume={142},
  number={2},
  pages={021004},
  year={2020},
  publisher={American Society of Mechanical Engineers}
}

@article{goh2024beyond,
  title={Beyond the stable handling limits: nonlinear model predictive control for highly transient autonomous drifting},
  author={Goh, Jonathan YM and Thompson, Michael and Dallas, James and Balachandran, Avinash},
  journal={Vehicle System Dynamics},
  volume={62},
  number={10},
  pages={2590--2613},
  year={2024},
  publisher={Taylor \& Francis}
}

@article{zarrouki2024r2nmpc,
  title={R$^2$nmpc: A real-time reduced robustified nonlinear model predictive control with ellipsoidal uncertainty sets for autonomous vehicle motion control},
  author={Zarrouki, Baha and Nunes, Joao and Betz, Johannes},
  journal={IFAC-PapersOnLine},
  volume={58},
  number={18},
  pages={309--316},
  year={2024},
  publisher={Elsevier}
}

@INPROCEEDINGS{10801876,
  author={Zarrouki, Baha and Wang, Chenyang and Betz, Johannes},
  booktitle={2024 IEEE/RSJ International Conference on Intelligent Robots and Systems (IROS)}, 
  title={Adaptive Stochastic Nonlinear Model Predictive Control with Look-ahead Deep Reinforcement Learning for Autonomous Vehicle Motion Control}, 
  year={2024},
  volume={},
  number={},
  pages={12726-12733},
  doi={10.1109/IROS58592.2024.10801876}}

@INPROCEEDINGS{7963824,
  author={Laurense, Vincent A. and Goh, Jonathan Y. and Gerdes, J. Christian},
  booktitle={2017 American Control Conference (ACC)}, 
  title={Path-tracking for autonomous vehicles at the limit of friction}, 
  year={2017},
  volume={},
  number={},
  pages={5586-5591},
  doi={10.23919/ACC.2017.7963824}}

@INPROCEEDINGS{9147713,
  author={Wurts, John and Dallas, James and Stein, Jeffrey L. and Ersal, Tulga},
  booktitle={2020 American Control Conference (ACC)}, 
  title={Adaptive Nonlinear Model Predictive Control for Collision Imminent Steering with Uncertain Coefficient of Friction}, 
  year={2020},
  volume={},
  number={},
  pages={4856-4861},
  doi={10.23919/ACC45564.2020.9147713}}

@INPROCEEDINGS{9867523,
  author={Vaskov, Sean and Quirynen, Rien and Menner, Marcel and Berntorp, Karl},
  booktitle={2022 American Control Conference (ACC)}, 
  title={Friction-Adaptive Stochastic Predictive Control for Trajectory Tracking of Autonomous Vehicles}, 
  year={2022},
  volume={},
  number={},
  pages={1970-1975},
  doi={10.23919/ACC53348.2022.9867523}}

@ARTICLE{11146591,
  author={Davydov, Alexander and Djeumou, Franck and Greiff, Marcus and Suminaka, Makoto and Thompson, Michael and Subosits, John and Lew, Thomas},
  journal={IEEE Robotics and Automation Letters}, 
  title={First, Learn What You Don't Know: Active Information Gathering for Driving at the Limits of Handling}, 
  year={2025},
  volume={10},
  number={11},
  pages={11379-11386},
  doi={10.1109/LRA.2025.3605097}}

@ARTICLE{10551497,
  author={Liu, Yingqiang and Chen, Zheng and Yao, Bin},
  journal={IEEE/ASME Transactions on Mechatronics}, 
  title={Online Optimization-Based Time-Optimal Adaptive Robust Control of Linear Motors With Input and State Constraints}, 
  year={2024},
  volume={29},
  number={4},
  pages={3157-3165},
  doi={10.1109/TMECH.2024.3404821}}

@misc{djeumou2023autonomousdrifting3minutes,
      title={Autonomous Drifting with 3 Minutes of Data via Learned Tire Models}, 
      author={Franck Djeumou and Jonathan Y. M. Goh and Ufuk Topcu and Avinash Balachandran},
      year={2023},
      eprint={2306.06330},
      archivePrefix={arXiv},
      primaryClass={eess.SY},
      url={https://arxiv.org/abs/2306.06330}, 
}

@article{spielberg2019neural,
  title={Neural network vehicle models for high-performance automated driving},
  author={Spielberg, Nathan A and Brown, Matthew and Kapania, Nitin R and Kegelman, John C and Gerdes, J Christian},
  journal={Science robotics},
  volume={4},
  number={28},
  pages={eaaw1975},
  year={2019},
  publisher={American Association for the Advancement of Science}
}

@article{yao1997adaptive,
  title={Adaptive robust control of SISO nonlinear systems in a semi-strict feedback form},
  author={Yao, Bin and Tomizuka, Masayoshi},
  journal={Automatica},
  volume={33},
  number={5},
  pages={893--900},
  year={1997},
  publisher={Elsevier}
}

@article{liu2025time,
  title={Time-optimal precision contouring control of biaxial gantry with online trajectory planning considering jerk constraint},
  author={Liu, Yingqiang and Lyu, Litong and Chen, Zheng and Yao, Bin},
  journal={IEEE Transactions on Industrial Electronics},
  year={2025},
  publisher={IEEE}
}

@article{yao2002indirect,
  title={Indirect adaptive robust control of SISO nonlinear systems in semi-strict feedback forms},
  author={Yao, Bin and Palmer, Andrew},
  journal={IFAC Proceedings Volumes},
  volume={35},
  number={1},
  pages={397--402},
  year={2002},
  publisher={Elsevier}
}

@INPROCEEDINGS{1243991,
  author={Bin Yao},
  booktitle={Proceedings of the 2003 American Control Conference, 2003.}, 
  title={Integrated direct/indirect adaptive robust control of SISO nonlinear systems in semi-strict feedback form}, 
  year={2003},
  volume={4},
  number={},
  pages={3020-3025 vol.4},
  doi={10.1109/ACC.2003.1243991}}

@misc{yu2025spatialenvelopempchigh,
      title={Spatial Envelope MPC: High Performance Driving without a Reference}, 
      author={Siyuan Yu and Congkai Shen and Yufei Xi and James Dallas and Michael Thompson and John Subosits and Hiroshi Yasuda and Tulga Ersal},
      year={2025},
      eprint={2509.18506},
      archivePrefix={arXiv},
      primaryClass={cs.RO},
      url={https://arxiv.org/abs/2509.18506}, 
}

@misc{verschueren2020acadosmodularopensourceframework,
      title={acados: a modular open-source framework for fast embedded optimal control}, 
      author={Robin Verschueren and Gianluca Frison and Dimitris Kouzoupis and Jonathan Frey and Niels van Duijkeren and Andrea Zanelli and Branimir Novoselnik and Thivaharan Albin and Rien Quirynen and Moritz Diehl},
      year={2020},
      eprint={1910.13753},
      archivePrefix={arXiv},
      primaryClass={math.OC},
      url={https://arxiv.org/abs/1910.13753}, 
}

@misc{zhou2025hierarchicaltestplatformvision,
      title={A Hierarchical Test Platform for Vision Language Model (VLM)-Integrated Real-World Autonomous Driving}, 
      author={Yupeng Zhou and Can Cui and Juntong Peng and Zichong Yang and Juanwu Lu and Jitesh H Panchal and Bin Yao and Ziran Wang},
      year={2025},
      eprint={2506.14100},
      archivePrefix={arXiv},
      primaryClass={cs.RO},
      url={https://arxiv.org/abs/2506.14100}, 
}
\bibliographystyle{IEEEtran}

\end{document}